\documentclass[10pt, aps,prl,twocolumn,showpacs,superscriptaddress,nofootinbib,preprintnumbers]{revtex4-2}
\usepackage{array}[=2016-10-06]
\usepackage{amsmath, graphicx}
\usepackage{xcolor}
\usepackage[colorlinks,linkcolor=blue,anchorcolor=blue,citecolor=blue,urlcolor=blue,]{hyperref}
\usepackage{orcidlink}
\usepackage{booktabs}
\usepackage{lineno}
\usepackage{float}
\usepackage{algorithm}
\usepackage{algpseudocode}
\usepackage{multirow}
\usepackage{diagbox}
\usepackage[caption=false]{subfig}
\usepackage{comment}
\usepackage[normalem]{ulem}
\usepackage{cancel}
\newcommand{\dd}{d}

\newcommand{\rs}{r_{\rm S}}
\newcommand{\ptot}{p_{\rm tot}}
\newcommand{\mdotout}{\dot{M}_{\rm out}}
\newcommand{\qdm}{q^{\chi e}}

\begin{document}

\title{Black hole accretion flows as dark matter bolometers} 

\author{Ran Ding}
\email{dingran@mail.nankai.edu.cn}
\affiliation{International Centre for Theoretical Physics Asia-Pacific (ICTP-AP),
University of Chinese Academy of Sciences (UCAS), Beijing, China}

\author{Ningqiang Song}
\email{songnq@itp.ac.cn}
\affiliation{Institute of Theoretical Physics, Chinese Academy of Sciences, Beijing, 100190, China}

\author{Cong Zhang}
\email{zhangcong@itp.ac.cn}
\affiliation{Institute of Theoretical Physics, Chinese Academy of Sciences, Beijing, 100190, China}

\date{\today}

\begin{abstract}

\end{abstract}

\begin{abstract}
Astrophysical systems provide powerful probes of dark matter interactions, complementing direct-detection searches. We propose using accretion flows of the supermassive black hole as a bolometer for light DM. The kinetic energy of sub-MeV dark matter in the black hole gravitational potential can remain below the thermal energy of the plasma electrons. The colder dark matter population can therefore absorb energy through elastic scattering, cooling the electrons and modifying their emission spectrum. We integrate dark matter cooling into the advection-dominated accretion flow around M87* and solve the full set of thermodynamic equations to obtain the resulting temperature profile and radio-to-millimeter emission.
By comparing the predicted spectrum with observations accompanying the Event Horizon Telescope campaign, we use, for the first time, the thermal energy balance of the hot accretion flow to constrain DM–electron scattering, obtaining stringent limits on keV-to-MeV DM for both heavy and light mediators. 
\end{abstract}

\maketitle
% \tableofcontents

%\noindent\textbf{\textit{Introduction}} -- 
\section{Introduction}

Cosmological and astrophysical observations have established the abundance of dark matter (DM)~\cite{Planck2018}, while leaving its particle mass and interactions undetermined.
Recent searches for weakly interacting massive particles (WIMPs) with PandaX-4T~\cite{Bo2025PandaX4T,Zhang2025PandaXLight}, XENONnT~\cite{XENON:2025WIMP,XENON:2026LightDM}, and LUX-ZEPLIN (LZ)~\cite{LZ:2024psa,LZ:2025light,LZ:2026extended} have improved sensitivity across a broad mass range, but have not confirmed a WIMP detection.
These results make light DM candidates particularly interesting and motivate searches that reach below the traditional WIMP mass range~\cite{Essig:2011nj,Chu:2011be,Essig:2015cda,Dvorkin:2019zdi,Boehm:2003hm,Lin:2011gj,Hochberg:2014dra,DAgnolo:2019zkf,Kahn:2021ttr}. Testing such light particles in conventional DM direct detection experiments is difficult because the recoil energies fall below detector thresholds as the mass decreases. In particular, sub-MeV DM is generally not energetic enough to induce ionization in noble liquids or semiconductors~\cite{Trickle:2019nya}.
New technology and experimental design that explore low-energy excitations are therefore required~\cite{Hochberg2016,Schutz2016,Hochberg2018,QROCODILE:2024nqm,Li:2025hqk}.

On the other hand, astrophysical systems provide complementary probes of light DM with the same interactions. Some astrophysical sources can reveal DM scattering through the energy loss in such systems. When cosmic rays transfer energy to ambient DM, the additional cooling changes their energy distribution and the radiation they produce.
This effect has been used to constrain various DM models~\cite{Cappiello2019,Ambrosone2023,Herrera2024,Gustafson2025,Mishra2025,Li2025,Chun2026,Hussein2025,Kantzas2026,Gorchtein2010,Cermeno2022}. 
However, the interpretation of CR-cooling bounds depends on the assumed
particle injection and transport, as well as conventional energy losses
~\cite{Cappiello2019,Ambrosone2023,Herrera2024}.

\begin{figure}[!t]
    \centering
    \includegraphics[width=\columnwidth]{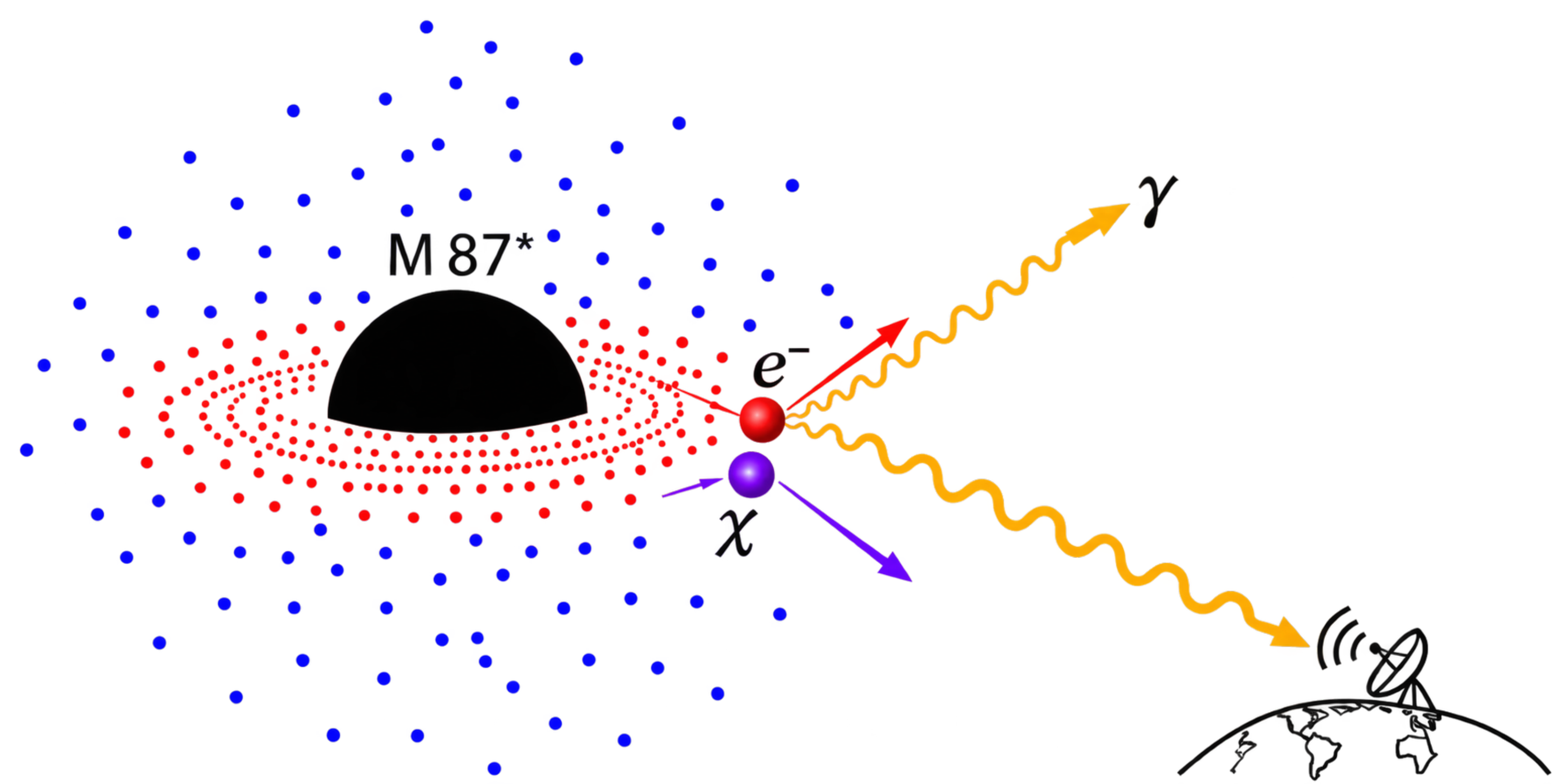}
    \caption{Schematic illustration of the mechanism in this work. DM in the halo (blue) around a supermassive black hole scatters off and cools the electrons in the accretion flow (red), which alters the emission spectrum from the accretion flow. }
    \label{fig:illustration}
\end{figure}

The extensively studied accretion flow around the supermassive black hole (SMBH) M87* offers a way to test dark matter (DM) scattering, since such scattering would alter the thermal state of the plasma and hence its broadband emission. The 2017 Event Horizon Telescope (EHT) campaign combined the first horizon-scale image with an extensive quasi-simultaneous multi-wavelength campaign from radio to gamma rays, yielding the most complete simultaneous spectrum of the active nucleus to date~\cite{Algaba2021}.
In an advection-dominated accretion flow (ADAF), the electron temperature is set by the balance of direct heating, Coulomb energy exchange with ions, advection, and radiative cooling~\cite{NarayanYi1995}. An additional cooling channel would therefore modify the thermal emission, whose contribution to the observed spectrum can be modeled jointly with the jet~\cite{Nemmen2014}.

DM could provide such a cooling channel around a SMBH, where the DM density is expected to be strongly enhanced by the adiabatic growth of the SMBH~\cite{Quinlan:1994ed,GondoloSilk1999,Ullio:2001fb,Gnedin:2003rj,Sadeghian2013,Lacroix:2016qpq}. The characteristic kinetic energy of a DM particle in the vicinity of the BH is of order $K_\chi\simeq m_\chi r_g/(2r)$, where $r_g\equiv GM_{\rm BH}$ is the gravitational radius of the BH (unless otherwise stated, we use natural units $\hbar=c=k_B=1$ throughout this work.). At the radii relevant to EHT emission, $r\sim5r_g$, this gives $K_\chi\simeq0.1$~MeV for a typical MeV-scale DM particle. In contrast, the plasma electrons have energies of a few MeV~\cite{EHTVIII2021}. Sub-MeV DM carrying less kinetic energy can therefore cool the electrons efficiently through scattering, as illustrated in Fig.~\ref{fig:illustration}.

In this \textit{Letter}, we incorporate DM–electron scattering into the ADAF equations and solve for the resulting electron temperature profile and the spectral energy distribution (SED) of the electron emission, focusing on the radio-to-submillimeter range, where this component is best constrained. For the first time, we exploit the thermal energy balance of the hot accretion flow, together with the 2017 multi-wavelength data~\cite{Algaba2021}, to set stringent constraints on light DM. These constraints reach DM masses well below those probed by direct-detection experiments, excluding cross sections of about $10^{-38}$~cm$^2$ for heavy-mediator models and $10^{-37}$~cm$^2$ for light-mediator models at $m_\chi=1$~keV. We also obtain limits on dark-photon kinetic mixing that improve upon existing bounds by up to about an order of magnitude.

\section{Modeling of the accretion flow and SED}

M87* is a low-luminosity black hole system whose bolometric luminosity lies many orders of magnitude below the Eddington limit. In this regime, its spectrum lacks the characteristic signatures of optically thick accretion, and only a small fraction of the rest-mass energy of accreted matter is converted into radiation. Such low-luminosity systems are instead naturally described by a geometrically thick, optically thin accretion flow composed of a two-temperature plasma consisting of electrons and ions, in which most of the dissipated energy is advected rather than promptly radiated---namely, an ADAF~\cite{Narayan:1994is,Narayan:1994et,Yuan:2000rj, Yuan:2003dc, Yuan:2014gma,Nemmen:2013mya}. Synchrotron radiation from this hot plasma also provides a natural origin for the millimeter emission observed from the near-horizon region of M87$^*$~\cite{Feng:2016gif,EventHorizonTelescope:2019pgp}. 

\begin{figure}[!htb]
    \centering
    \includegraphics[width=\columnwidth]{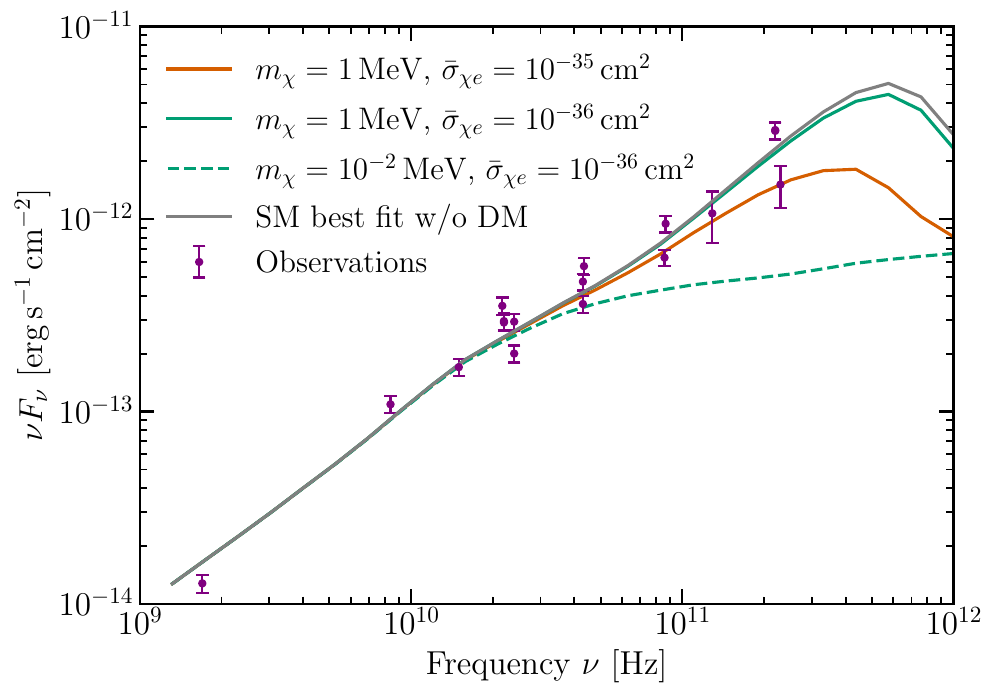}
    \caption{Comparison of the predicted M87$^*$ SEDs with observational data. The data points with error bars show measurements obtained with the EVN~\cite{Cheung:2007wp,Chang:2010xe,Giroletti:2012ge,Hada:2015vka}, HSA, VERA~\cite{minh2003new}, EAVN~\cite{qain2016frontiers,Asada:2017hky,an2018capabilities}, KVN~\cite{kim2018long}, VLBA~\cite{CraigWalker:2018vam}, GMVA~\cite{Boccardi:2017npi}, ALMA~\cite{matthews2018alma}, SMA~\cite{2007ASPC..375..234G}, and EHT~\cite{Algaba2021} (compiled in~\cite{Algaba2021}), carried out quasi-simultaneously with the 2017 EHT campaign. The gray line shows the best-fit baseline spectrum obtained from Standard Model (SM) processes, while the colored lines include the effect of DM cooling by scattering with electrons. We assume a heavy mediator for the DM models.  Results for a light mediator are qualitatively similar.}
\label{fig:spectrum-comparison}
\end{figure}

The ADAF framework explicitly tracks the ion and electron energy budgets through Eq.~\eqref{eq:adaf}, which incorporates the direct heating rate $q^+$ from viscous dissipation, the Coulomb energy exchange $q^{ie}$, the advection rate $q^{\text{adv}}$, and the radiative cooling rate $q^{-}$ of the electrons. A fraction $\delta$ of the viscously dissipated energy is deposited directly into the electrons, while the remaining fraction $1-\delta$ heats the ions. Coulomb coupling is insufficient to maintain a common temperature, so the electrons and ions are characterized by separate temperatures $T_e$ and $T_i$. Schematically, the radial entropy advection rates can be written as~\cite{Yuan:2000rj,Yuan:2003dc,Yuan:2014gma}
\begin{subequations}\label{eq:adaf}
\begin{align}
q^{\mathrm{adv},i}
&=\rho v_r T_i\frac{ds_i}{dr}
=(1-\delta)q^+-q^{ie},
\label{eq:ion-energy}\\
q^{\mathrm{adv},e}
&=\rho v_r T_e\frac{ds_e}{dr}
=\delta q^++q^{ie}-q^-,
\label{eq:electron-energy}
\end{align}
\end{subequations}
where $\rho$ is the gas mass density, $v_r<0$ is the common radial inflow velocity, and $s_i$ and $s_e$ are the ion and electron contributions to the entropy per unit total gas mass. The emission near the first peak of the spectrum arises mainly from synchrotron radiation by thermal electrons in the plasma near the black hole. This radiation is highly self-absorbed and extremely sensitive to the electron temperature, scaling as $\nu F_\nu\propto T_e^7$, where $F_\nu$ is the SED~\cite{Yuan:2014gma}.

To solve the ADAF equations and compute the resulting SED, we use the \texttt{riaf-sed} code~\cite{Nemmen:2013mya,Nemmen:RIAFSED,Yuan:2004iw,Nemmen_2014,Yuan:2006nd}. The code models the ADAF semi-analytically as a stationary accretion flow, adopting the $\alpha$-viscosity prescription and a pseudo-Newtonian gravitational potential, and treats radiative transfer self-consistently, including synchrotron, bremsstrahlung, and inverse-Compton emission. We perform a Markov chain Monte Carlo (MCMC) analysis, fitting the model to the observed SED of M87$^*$, with the outer accretion rate $\dot{M}_{\mathrm{out}}/\dot{M}_{\mathrm{Edd}}$, the electron-heating fraction $\delta$, the outflow index $s$, and the jet contribution to the SED as free parameters. As shown in Fig.~\ref{fig:spectrum-comparison}, the best-fit baseline spectrum, hereafter referred to as the SM best-fit, reproduces the broadband SED well over the frequency range $1.7$--$230\,\mathrm{GHz}$~\cite{EventHorizonTelescope:2021dvx}. The low-frequency emission is dominated by the jet, whereas the ADAF becomes increasingly important toward millimeter wavelengths and dominates in the $\sim$ 230 GHz band probed by the EHT. Further details on the flow structure of M87$^*$, the ADAF model, and the corresponding broadband SED are given in the Supplemental Material. We truncate the model SED at $10^{12}$Hz, above which it may be dominated by emission from non-thermal electrons~\cite{Yuan:2014gma,Yuan:2003dc,Bandyopadhyay:2019caa}.

\section{Dark matter-induced electron cooling}

As the SED is highly sensitive to the electron temperature, DM–electron scattering in the presence of ambient DM provides an additional cooling channel: it lowers the electron temperature and thereby modifies the SED. We quantify this effect through the volumetric DM-induced electron cooling rate,
\begin{equation}\label{eq:DMcoolingrate}
q^{\chi e}(r)=n_e(r)\left\langle \dot E_e(r)\right\rangle_{\rm th}\,,
\end{equation}
where $\langle\cdots\rangle_{\rm th}$ denotes the thermal average over the electron distribution function at temperature $T_e (r)$ (see the Supplemental Material). We define
the single-electron energy-loss rate from scattering with ambient DM 
particles as $\dot E_e\equiv-dE_e/dt$, which is positive for cooling,
\begin{equation}\label{eq:cooling}
\dot E_e(E_e^{},r)=\int f_\chi p^2 dp\int d\Omega\int \left(-\Delta E\right)\,d\sigma\,v_{\chi e}\,.
\end{equation}
Here $f_\chi(p,r)$ is the phase-space distribution function of DM particles, $d\sigma$ is the cross section for DM-electron elastic scattering, and $v_{\chi e}$ is the relative velocity between the incoming DM particle and the electron, while
$-\Delta E\equiv E_e^{}-E_{e,\rm final}^{}$ denotes the energy lost by the electron.

The innermost integral represents the cooling rate per electron per scattering event, for a given initial electron energy and incoming DM direction $\Omega$. In evaluating Eq.~\eqref{eq:cooling}, we assume the virialized DM velocity distribution to be isotropic, with magnitude $v_\chi(r)=(r_g/r)^{1/2}$.
Under this approximation, Eq.~\eqref{eq:cooling} can be recast as (see the Supplemental Material for details)
\begin{equation}\label{eq:coolingreduced}
\dot E_e(E_e,r)
=\frac{\rho_\chi(r)}{4\pi m_\chi}\int d\Omega\int \left(-\Delta E\right)
\frac{\overline{|\mathcal{M}|^2}}{2E_\chi\,2E_e}\,d\Pi_2\,,
\end{equation}
where $\overline{|\mathcal{M}|^{2}}$ is the squared DM--electron scattering matrix element, averaged over initial and summed over final spin states, and $d\Pi_2$ is the two-body Lorentz-invariant phase space; $E_e$ and $E_\chi$ denote the incident electron and DM energies. 

For the DM density profile $\rho_\chi(r)$, we adopt the spike model of Ref.~\cite{GondoloSilk1999} for the non-annihilating case, with the parametrization and normalization of Refs.~\cite{Lacroix:2015lxa,Lacroix:2016qpq,Chen:2024nua}:
\begin{equation}\label{eq:density}
\rho_\chi(r)=1.01\times10^{8}
\left(\frac{r}{1\,\mathrm{pc}}\right)^{-7/3}\,\mathrm{GeV\,cm^{-3}}\,,
\end{equation}
which applies over the range $r_{\rm crit}(\theta,a_J)\leq r<r_{\rm sp}$. Here, $r_{\rm crit}(\theta,a_J)$ defines the inner cutoff radius of the DM density profile~\cite{Hod:2013mgr}: for M87$^*$ with spin parameter $a_J=0.9375$, $r_{\rm crit}$ ranges from $1.5\,r_g$ in the equatorial plane ($\theta=\pi/2$) to $3.5\,r_g$ along the polar axis ($\theta=0$), and $r_{\rm sp}=3.87\times10^{5}\,r_g$ is the outer radius of the DM spike. This radial range covers the region relevant for the cooling of the inner accretion flow.

Although our framework applies to general DM--electron interactions, as a benchmark we consider DM charged under $U(1)_{A'}$, where the dark photon mixes with the SM photon with the kinetic mixing parameter $\epsilon$. With field redefinition, the relevant interaction Lagrangian is
\begin{equation}\label{eq:lagrangian}
\mathcal{L}\supset g_\chi\bar{\chi}\gamma^\mu\chi A^\prime_\mu+\epsilon g_{\rm em}\bar{e}\gamma^\mu e A^\prime_\mu\,,
\end{equation}
where $g_\chi$ and $g_{\rm em}$ are the dark coupling and electromagnetic coupling. The kinematics for DM-electron scattering and the associated cooling rate are detailed in the Supplemental Material. 

We then incorporate DM cooling into the evolution of the accretion flow by adding $-q^{\chi e}(r)$, defined in Eq.~\eqref{eq:DMcoolingrate}, to the right-hand side of Eq.~\eqref{eq:electron-energy}. Since the squared matrix element contains an overall factor of $g_\chi^2 \epsilon^2 g_{\rm em}^2$, it is convenient to parametrize the interaction strength in terms of the reference DM--electron scattering cross section $\bar{\sigma}_{\chi e}$ conventionally used in direct-detection studies,
\begin{equation}
\bar{\sigma}_{\chi e}\equiv\frac{\mu_{\chi e}^2}{\pi}\frac{g_\chi^2\epsilon^2g_{\rm em}^2}{\left(m_{A'}^2+\alpha_{\rm em}^2m_e^2\right)^2}\,,
\end{equation}
where $\mu_{\chi e}=m_\chi m_e/(m_\chi+m_e)$ is the DM--electron reduced mass and $\alpha_{\rm em}$ is the electromagnetic fine-structure constant. We use the \texttt{riaf-DM} code\footnote{The code is publicly available on GitHub at \url{https://github.com/CongZhang-123/RIAF-DM}.}, which we developed based on the \texttt{riaf-sed} code, to include the effect of DM cooling consistently and to compute the modified SEDs. All other parameters are kept fixed at the SM best-fit values.

The colored curves in Fig.~\ref{fig:spectrum-comparison} show the SEDs including DM--electron scattering. Relative to the SM best fit, they are suppressed at millimeter wavelengths. For sufficiently large cross sections, the SED converges to the jet-only case. We fix the jet contribution and neglect its cooling through scattering with DM; including this effect would further suppress the SED in the same frequency range. Lowering the DM mass also suppresses the SEDs, because the DM number density increases and the DM kinetic energy decreases relative to that of the thermal electrons. Both effects enhance the DM-induced cooling rate, which lowers the electron temperature and reduces the radiative output of the accretion flow. As shown in the Supplemental Material, DM cooling mainly lowers the electron temperature in the inner accretion flow, within $r\lesssim30\,r_g$.

\section{Constraints on dark matter}
\begin{figure*}[!t]
    \centering
    \includegraphics[width=\columnwidth]{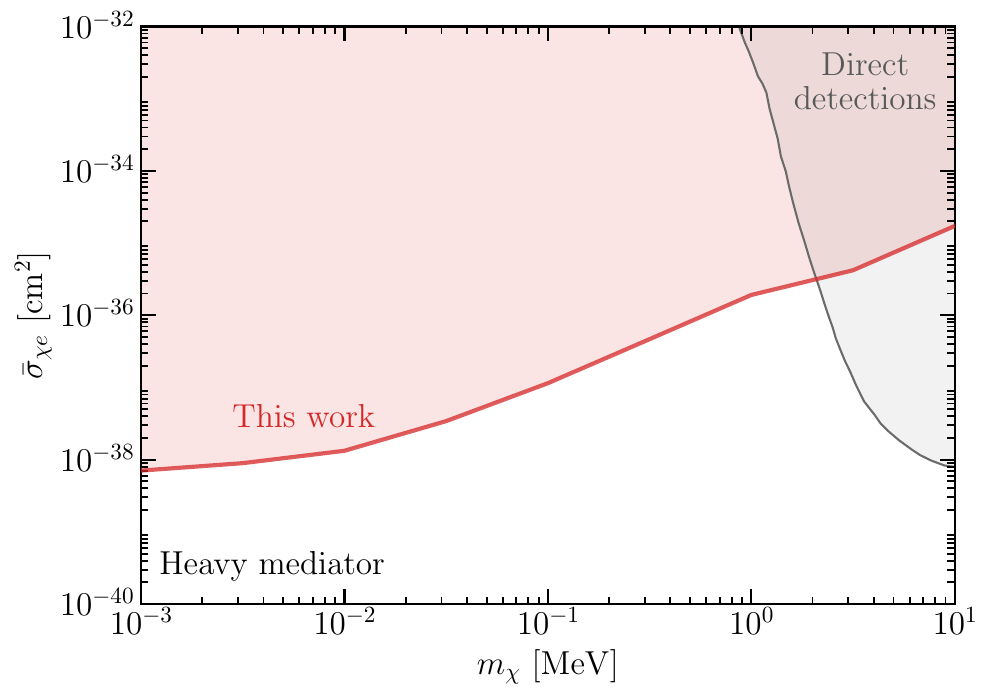}
    \vspace{0.5em}
    \includegraphics[width=\columnwidth]{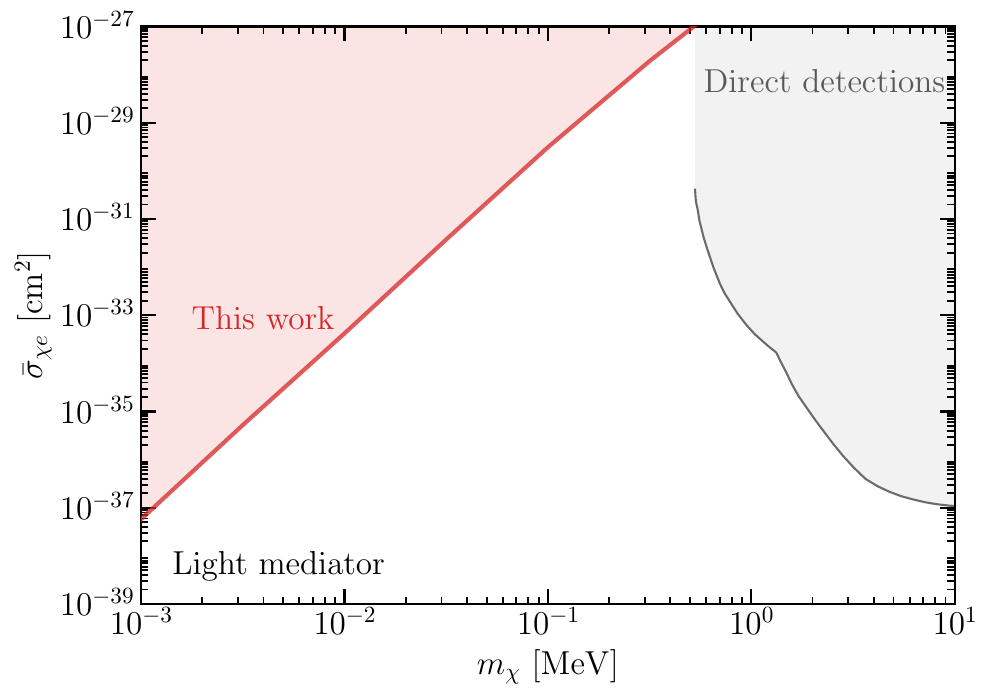}
    \caption{90\% C.L. limits on the reference DM--electron scattering cross section $\bar{\sigma}_{\chi e}$ as a function of the DM mass $m_\chi$ for a heavy mediator (left) and a light mediator (right). The red regions show the constraints from DM cooling of the accretion flow obtained in this work, and the gray regions display the constraints from existing direct detection experiments~\cite{SENSEI:2023zdf,DAMIC-M:2025luv}.}
    \label{fig:sigmadm-limit-vs-mchi}
\end{figure*}

We then vary the DM mass and the reference scattering cross section $\bar{\sigma}_{\chi e}$, compare the predicted SED with the observed one in a Gaussian likelihood analysis, and derive the 90\% confidence level (C.L.) upper limit on $\bar{\sigma}_{\chi e}$.

We present the constraints obtained from DM cooling of the accretion flow in Fig.~\ref{fig:sigmadm-limit-vs-mchi}  for the cases of a light and a heavy mediator. For a massless mediator, the DM cooling rate is infrared divergent and should be regulated by in-medium polarizations according to the properties of the plasma~\cite{DeRocco2022,BraatenThoma1991}. We therefore choose the mediator mass $m_{A'}=0.1$~eV, well above the plasma frequency in the inner accretion flow. For halo DM in terrestrial direct-detection experiments, the momentum transfer is much larger than $m_{A'}$ over the keV--MeV DM mass range, so the mediator is effectively light. We have explicitly verified that the DM cooling constraints vary negligibly for $m_{A'}\lesssim1$~keV.

The constraints become more stringent at lower DM masses. Towards $m_\chi\sim1$~keV, the limits on the reference cross section are comparable for light and heavy mediators. As the DM mass increases, the limit for a light mediator weakens much faster than that for a heavy mediator. This behavior can be understood qualitatively. Define the mediator form factor $F_{\rm med}^2=
\left[m_{A'}^2+(\alpha_{\rm em}m_e)^2\right]^2/
\left(m_{A'}^2-t_M\right)^2$, assuming isotropic scattering in the center-of-mass frame and DM is at rest; after thermal and angular averaging, the Mandelstam variable scales as $t_M\propto m_\chi^2T_e^2/m_e^2$ for light DM with $m_\chi\ll m_e$. The form factor therefore falls rapidly with increasing DM mass for a light mediator, whereas $F_{\rm med}^2\simeq1$ for a heavy mediator regardless of $m_\chi$. For completeness, we also show the limits for different mediator masses in the Supplemental Material.

In Fig.~\ref{fig:sigmadm-limit-vs-mchi}, we also show the existing constraints from direct detection experiments. In the relevant parameter range, SENSEI~\cite{SENSEI:2023zdf} and DAMIC-M~\cite{DAMIC-M:2025luv} provide the strongest constraints. For a heavy mediator, the DM-cooling limits surpass those from direct detection at $m_\chi\lesssim2$~MeV. For both heavy mediator and light mediator models, DM cooling sets stringent limits in the sub-MeV DM mass range, where direct-detection signals typically fall below experimental thresholds. We also note limits from DM that is accelerated and reflected by solar electrons and detected in terrestrial direct-detection experiments~\cite{An:2017ojc,An:2021qdl,Emken:2024nox,PandaX:2024syk}; these are comparable to or weaker than the DM-cooling limits at $m_\chi\gtrsim0.01$~MeV and are subject to different systematics. For direct detection and solar reflection alike, the limits saturate at large cross sections, where frequent scattering in the overburden decelerates DM before it reaches the underground detector~\cite{Emken:2019tni}. In contrast, the same interaction enhances DM cooling, so the DM-cooling limits are not bounded from above.

\begin{figure}[!htb]
    \centering
    \includegraphics[width=\columnwidth]{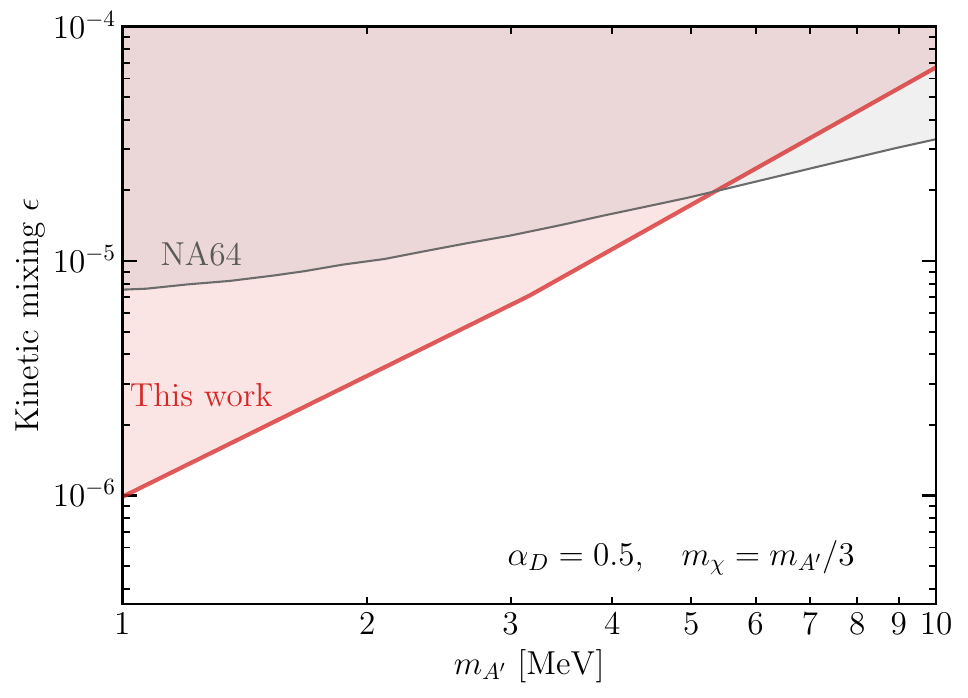}
    \caption{90\% C.L. limits on the kinetic mixing
    parameter $\epsilon$ as a function of the dark photon mass $m_{A'}$. The red region shows the constraints from DM cooling of the accretion flow obtained in this work, and the gray region depicts the constraints from the NA64 experiment~\cite{Andreev:2021fzd}. We choose
    $\alpha_D=g_\chi^2/4\pi=0.5$ and $m_\chi=m_{A'}/3$ for a consistent comparison.}
    \label{fig:epsilon-vs-ma-na64}
\end{figure}

With the model assumed in Eq.~\eqref{eq:lagrangian}, we can also constrain the dark photon mediator, for which the mediator mass and the kinetic mixing parameter affect the DM cooling rate. We fix the relation between $m_\chi$ and $m_{A'}$, and the dark coupling to match the NA64 benchmark values~\cite{Andreev:2021fzd}, where the dark photon decaying invisibly will contribute to missing energy in the detector. As shown in Fig.~\ref{fig:epsilon-vs-ma-na64}, the DM-cooling limits become more stringent as $m_{A'}$ decreases, exceeding the NA64 limits for $m_{A'}\lesssim5$~MeV. At $m_{A'}=1$~MeV, the cooling limits improve upon those from NA64 by about an order of magnitude.

\section{Conclusions and prospects}

Light DM can efficiently cool the electrons in the inner accretion flow around a SMBH, thereby altering their emission spectrum. Taking the M87$^*$ SED as an example, we set stringent constraints on sub-MeV DM and on invisibly decaying MeV-scale dark photon by confronting the predicted emission spectrum with observations.

Although we adopt the dark photon-mediated model as a benchmark, the energy-balance framework can be extended to other mediators, to transitions between different DM states, and to the production or absorption of dark-sector particles, once the corresponding collision terms and transport are included. Heavier DM may instead heat the plasma by transferring the kinetic energy gained in the black-hole potential. We leave these extensions to future work.

% DM coupling predominantly to nucleons would exchange energy directly with the ions. However, Coulomb energy exchange between ions and electrons is inefficient, and our numerical tests show that the resulting change in the electron temperature is negligible for the parameters studied.  {\color{red}Nucleon-coupled DM therefore leaves the SED essentially unchanged in our setup.}

The same framework can be applied to other SMBHs that host hot, radiatively inefficient accretion flows with well-measured nuclear spectra, such as Sgr~A$^*$~\cite{EventHorizonTelescope:2022urf}. Sgr~A$^*$ provides a complementary laboratory to M87$^*$: its black hole mass is about three orders of magnitude smaller and its accretion flow is more variable, so the same DM-cooling analysis would probe a different combination of plasma temperature, density, and DM environment. Studying such sources could provide complementary constraints and help assess the impact of uncertainties in the accretion flow and DM distribution.

Future studies could incorporate DM--electron energy exchange into two-temperature general relativistic magnetohydrodynamic (GRMHD) simulations~\cite{Ressler:2015ipa}. By lowering the electron temperature, DM cooling modifies the synchrotron emissivity and absorption, and can change the Faraday rotation, thereby altering the brightness distribution and polarization of the near-horizon emission. Polarized radiative transfer would connect these changes to EHT observations of the photon ring, allowing joint analyses of the spectrum and interferometric data to constrain DM scattering while accounting for uncertainties in electron heating and accretion physics, potentially setting stronger and more robust constraints on DM and dark sector particles~\cite{EventHorizonTelescope:2019pgp,EHTVIII2021}.  Next-generation EHT upgrades with extended baselines would further improve the sensitivity of such joint analyses.

\section{Acknowledgements}
We thank Yuxin Liu for valuable contributions in the early stages of this work. R.D. thanks Huai-Ke Guo for hosting him at ICTP-AP, UCAS. We also acknowledge the high-performance computing resources provided by the Institute of Theoretical Physics, Chinese Academy of Sciences (ITP-CAS), for the MCMC calculations. N.S. is supported by the National Natural Science Foundation of China (NSFC) Project Nos. 12475110, 12347105, 12441504, and 12447101. C.Z. is supported by the NSFC Project Nos. 12605181 and 12447101.

\bibliography{ref}

\clearpage

\clearpage
\onecolumngrid

\begin{center}
{\large\bfseries Supplemental Material}
\end{center}
%\vspace{1em}

\setcounter{section}{0}
\setcounter{subsection}{0}
\setcounter{equation}{0}
\setcounter{figure}{0}
\setcounter{table}{0}

\setcounter{secnumdepth}{2}

\renewcommand{\thesection}{S\arabic{section}}
\renewcommand{\thesubsection}{\thesection.\arabic{subsection}}
\renewcommand{\theequation}{S\arabic{equation}}
\renewcommand{\thefigure}{S\arabic{figure}}
\renewcommand{\thetable}{S\arabic{table}}

\renewcommand{\theequation}{S\arabic{equation}}
\renewcommand{\thefigure}{S\arabic{figure}}
\renewcommand{\thetable}{S\arabic{table}}
\renewcommand{\bibnumfmt}[1]{[#1]}
\renewcommand{\citenumfont}[1]{#1}

\setcounter{page}{1}

This Supplemental Material presents the accretion-flow equations and
baseline spectral fit, derives the relativistic DM--electron
energy-transfer rate, and shows the resulting temperature profiles and
constraints for different mediator masses.

\section{ADAF modeling of M87*: flow structure and spectrum}
\label{sup:sec-adaf}

\subsection{Dynamical and thermal equations}
The stationary, axisymmetric ADAF is described by the following
height-integrated equations
\citep{Yuan:2000rj,Yuan:2003dc,Yuan:2014gma,Nemmen:2013mya}:
\begin{subequations}\label{sup:eq-flow}
\begin{align}
 \dot M(r)&=4\pi\rho rH|v_r|
 =\mdotout\left(\frac{r}{r_{\rm out}}\right)^s,
 \label{sup:eq-mass}\\
 v_r\frac{\dd v_r}{\dd r}-\Omega^2r
 &=-\Omega_K^2r-\frac{1}{\rho}\frac{\dd\ptot}{\dd r},
 \label{sup:eq-radial}\\
 v_r\frac{\dd(\Omega r^2)}{\dd r}
 &=\frac{1}{\rho rH}\frac{\dd}{\dd r}
 \left(\nu_{\rm vis}\rho r^3H\frac{\dd\Omega}{\dd r}\right),
 \label{sup:eq-angular}\\
 q^{\mathrm{adv},i}
 &\equiv\rho v_r\left(\frac{\dd e_i}{\dd r}
 -\frac{p_i}{\rho^2}\frac{\dd\rho}{\dd r}\right)
 =(1-\delta)q^+-q^{ie},
 \label{sup:eq-ion}\\
 q^{\mathrm{adv},e}
 &\equiv\rho v_r\left(\frac{\dd e_e}{\dd r}
 -\frac{p_e}{\rho^2}\frac{\dd\rho}{\dd r}\right)
 =\delta q^++q^{ie}-q^- .
 \label{sup:eq-electron}
\end{align}
\end{subequations}
Eq.~\eqref{sup:eq-mass} describes mass conservation with an inwardly
decreasing accretion rate when $s>0$. Here $\rho$ is the gas mass density
and $v_r<0$ is the common radial inflow velocity. Vertical hydrostatic
balance gives the scale height $H\simeq c_s/\Omega_K$, where
$c_s^2=\ptot/\rho=(p_i+p_e)/\beta /\rho$ defines the isothermal sound speed, and $\Omega_K$ is the Kaplerian angular velocity. $p_i$
and $p_e$ are the ion and electron thermal pressures.
Eqs.~\eqref{sup:eq-radial} and \eqref{sup:eq-angular} describe
radial force balance and viscous angular-momentum transport;
$\Omega$ is the flow angular velocity and $\nu_{\rm vis}(=\alpha c_s H) $ is the
kinematic viscosity coefficient. 
% The Paczy\'nski--Wiita potential is given in the
% first paragraph of Sec.~2 of \citet{Yuan:2000rj}; it determines the
% Keplerian frequency through
% \[
%  \Phi(r)=-\frac{G\mbh}{r-\rs},\qquad
%  \Omega_K^2=\frac{1}{r}\frac{\dd\Phi}{\dd r}
%  =\frac{G\mbh}{r(r-\rs)^2},
% \]
% where $\rs=2\rg$ is denoted $r_g$ in that reference.
% We include the wind through $\dot M(r)$ without
% an additional torque term, following the approximation discussed in
% \citet{Yuan:2014gma}.
Eqs.~\eqref{sup:eq-ion} and \eqref{sup:eq-electron} are the ion and
electron energy equations of the main text, expressed in terms of the
specific internal energies $e_i$ and $e_e$ per unit gas mass. The pressure
terms account for compressional work. DM cooling is included by adding
$-\qdm$ to the right hand side of Eq.~\eqref{sup:eq-electron}.

\subsection{Baseline spectral fit}
\begin{table}[!htb]
\centering
\caption{Parameters of the baseline SED fit for M87$^*$. The outer accretion rate is $\dot M_{\rm out}\equiv\dot M(r_{\rm out})$, and the Eddington accretion rate is $\dot M_{\rm Edd}=10L_{\rm Edd}$. The outer temperatures are normalized to the virial temperature $T_{\rm vir}(r)=3.6\times10^{12}(r_s/r)\,\mathrm{K}$, with $r_s=2r_g$. The mass of M87$^*$ $M_{\rm BH}=6.5\times10^9\,M_\odot$~\cite{EventHorizonTelescope:2019dse,EventHorizonTelescope:2019ggy}, corresponding to $r_g=3.1\times10^{-4}\,\mathrm{pc}$.}
\label{sup:tab-parameters}
\small
\renewcommand{\arraystretch}{1.15}
\begin{tabular}{@{}p{0.18\textwidth}p{0.43\textwidth}p{0.08\textwidth}p{0.22\textwidth}@{}}
\toprule
Parameter & Physical meaning & Status & Prior range or value\\
\midrule
$\mdotout/\dot M_{\rm Edd}$ & Outer mass-accretion rate in Eddington units
& Varied & $[5.5\times10^{-5},5.5\times10^{-3}]$\\
$\delta$ & Fraction of viscous dissipation heating electrons directly
& Varied & $[10^{-3},10^{-1}]$\\
$s$ & Index of the inward mass-accretion profile & Varied & $[10^{-2},1]$\\
$j_{\rm factor}$ & Multiplicative normalization of the jet spectrum
& Varied & $[0.25,4]$\\
$\alpha$ & Viscosity parameter & Fixed & $0.3$\\
$\beta$ & Gas-to-total pressure ratio & Fixed & $0.9$\\
$\gamma$ &  Adiabatic index & Fixed & $1.5$\\
$r_{\rm out}/\rs$ & Outer boundary  of the ADAF & Fixed & $10^4$\\
$T_{i,\rm out}/T_{\rm vir}$ & Ion temperature at the outer boundary  & Fixed & $0.2$\\
$T_{e,\rm out}/T_{\rm vir}$ & Electron temperature at the outer boundary & Fixed & $0.19$\\
$\lambda_{\rm out}$ & Radial Mach number at the outer boundary $|v_r|/c_s$
& Fixed & $0.2$\\
\bottomrule
\end{tabular}
\end{table}
We fit the baseline spectrum with \textsc{riaf-sed}
\citep{Nemmen:RIAFSED}, varying the
parameters controlling the mass supply, outflow strength, and direct
electron heating, together with the jet normalization. The remaining
model parameters and outer boundary conditions are held fixed at the
benchmark values adopted from \citet{Nemmen:2013mya}.
Table~\ref{sup:tab-parameters} lists the varied parameters and their
ranges, as well as the fixed quantities. The jet spectral shape follows
\citet{Nemmen:2013mya}, with its amplitude rescaled by the
frequency-independent factor $j_{\rm factor}$.
We also reproduce their reference SED as a check of the spectral
calculation.

\begin{figure}[tbp]
\centering
\includegraphics[width=0.85\textwidth]{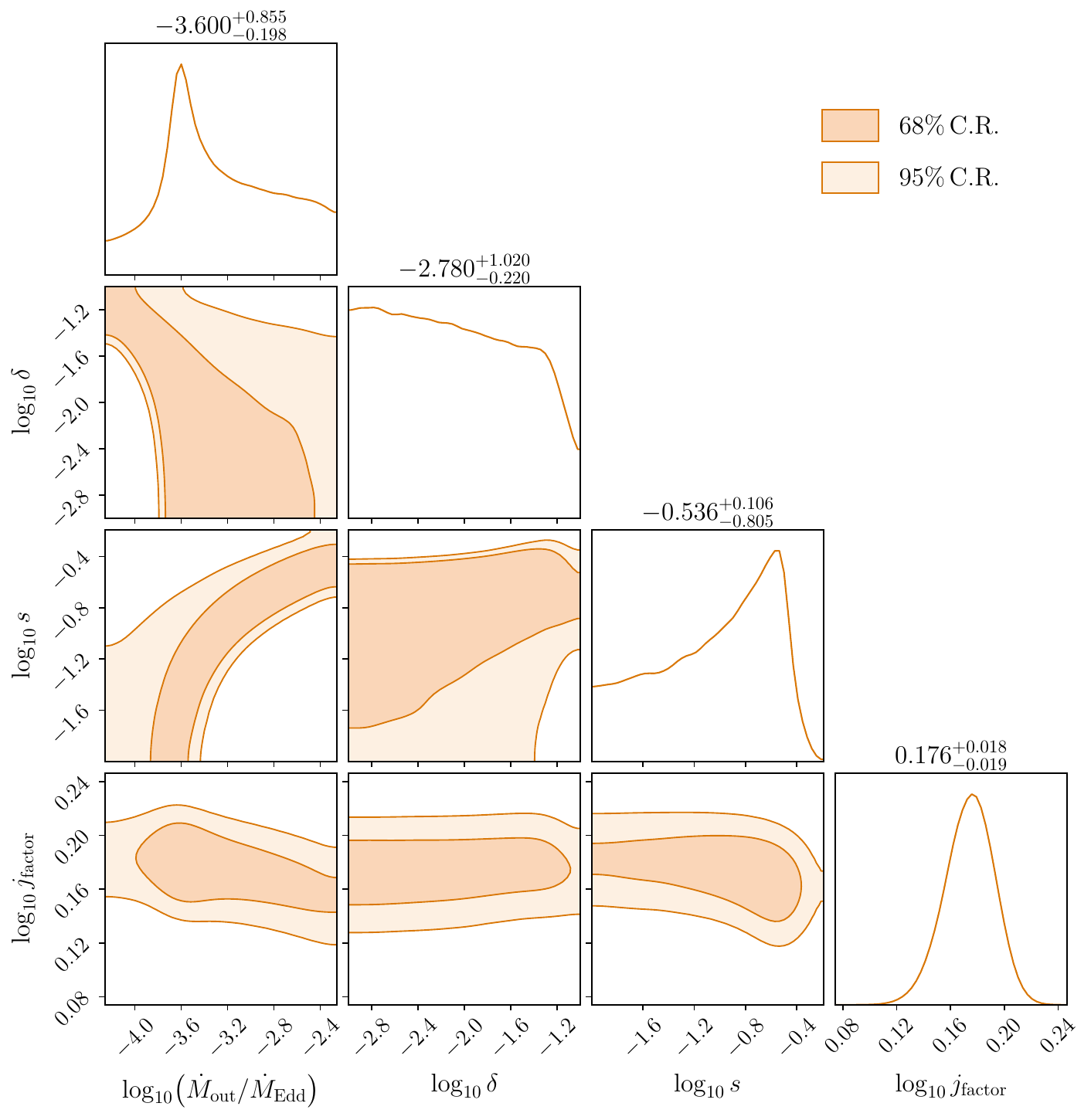}
\caption{Marginalized posterior distributions for the four parameters
of the baseline M87* SED fit. Diagonal panels show one-dimensional
posteriors; the joint contours enclose 68\% and 95\% credible regions.
Annotations give the 1-d marginalized best fit and 68\% range in
log space.}
\label{sup:fig-mcmc}
\end{figure}

Fig.~\ref{sup:fig-mcmc} shows the marginalized posterior distributions
and joint credible regions (C.R.). The global best-fit model has
$\chi^2_{\min}=67.8$ with ($\log_{10}(\dot{M}_{\mathrm{out}}/\dot{M}_{\mathrm{Edd}})$,
$\log_{10}\delta$, $\log_{10}s$,
$\log_{10}j_{\mathrm{factor}}$)=($-2.300$, $-2.224$, $-0.468$, $0.158$). It defines the fixed astrophysical baseline for
the subsequent DM analysis.  The model parameters predict the electron temperature $T_e=1.6\times10^{10}\,\mathrm{K}$, electron number density $n_e=1.7\times10^6\,\mathrm{cm}^{-3}$, and the disk magnetic field $B=20\,\mathrm{G}$ at $r=5r_g$, which are consistent
with those
estimated by the EHT Collaboration using the one-zone model~\cite{EHTVIII2021}.

\begin{figure}[tbp]
\centering
\includegraphics[width=0.6\textwidth]{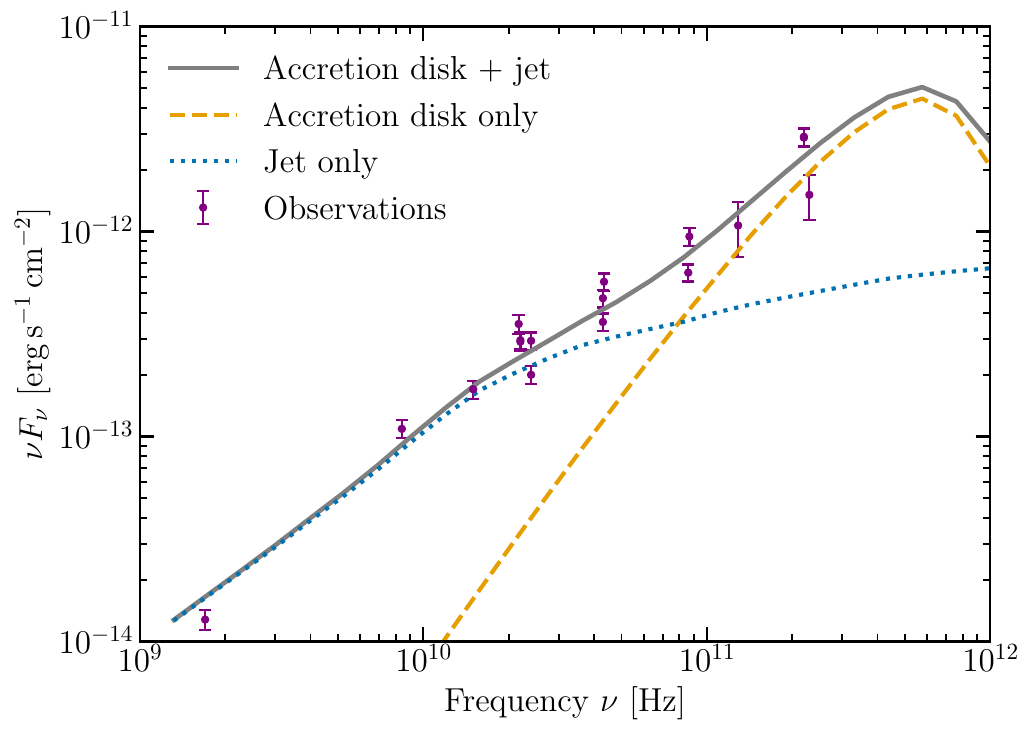}
\caption{Baseline best-fit SED of M87*, decomposed into the ADAF and jet components. Their sum is compared with the
radio-to-millimeter observations accompanying the 2017 EHT campaign
\citep{EventHorizonTelescope:2021dvx}.}
\label{sup:fig-baseline}
\end{figure}

Fig.~\ref{sup:fig-baseline} compares this spectrum with the
$1.7$--$230\,\mathrm{GHz}$ observations accompanying the 2017 EHT
campaign \citep{EventHorizonTelescope:2021dvx}. The decomposition shows
the transition from jet-dominated radio emission to the increasing
contribution of the accretion flow toward millimeter wavelengths.

\section{Relativistic dark matter--electron scattering and energy transfer}
\label{sup:scattering}

To evaluate the electron energy-loss rate, we consider the scattering
\begin{equation}
e^-(p_1)+\chi(p_2)\longrightarrow e^-(p_3)+\chi(p_4).
\label{sup:process}
\end{equation}
The initial momenta can be written as
\begin{equation}
\begin{aligned}
p_1^\mu&=(E_e,0,0,p_e),\\
p_2^\mu&=(E_\chi,p_\chi\sin\theta,0,p_\chi\cos\theta),
\end{aligned}
\label{sup:lab-momenta}
\end{equation}
where $p_e=\sqrt{E_e^2-m_e^2}$, $p_\chi=\sqrt{E_\chi^2-m_\chi^2}$,
and $\theta$ is the angle between the incident momenta.
The total energy and boost to the center-of-mass (CM) frame are
\begin{equation}
\begin{aligned}
E_{\rm tot}&=E_e+E_\chi,&
\boldsymbol\beta_{\rm CM}
&=(p_\chi\sin\theta,0,p_e+p_\chi\cos\theta)/E_{\rm tot},\\
\beta_{\rm CM}&=\sqrt{1-\frac{s_{\rm M}}{E_{\rm tot}^2}},&
\gamma_{\rm CM}&=\frac{E_{\rm tot}}{\sqrt{s_{\rm M}}},
\end{aligned}
\label{sup:cm-boost}
\end{equation}
with the Mandelstam variable
\begin{align}
s_{\rm M}=(p_1+p_2)^2
=m_e^2+m_\chi^2+2(E_eE_\chi-p_ep_\chi\cos\theta)\,.
\end{align}

Applying the Lorentz boost defined above, the momenta in the CM frame are
\begin{equation}
\begin{aligned}
p_1^{*\mu}&=\left(E_e^*,\,p_*^x,\,0,\,p_*^z\right),\\
p_2^{*\mu}&=\left(E_\chi^*,\,-p_*^x,\,0,\,-p_*^z\right),\\
p_3^{*\mu}&=(E_e^*,p_*\sin\theta_*\cos\phi_*,
                     p_*\sin\theta_*\sin\phi_*,p_*\cos\theta_*),\\
p_4^{*\mu}&=(E_\chi^*,-p_*\sin\theta_*\cos\phi_*,
                         -p_*\sin\theta_*\sin\phi_*,-p_*\cos\theta_*),
\end{aligned}
\label{sup:cm-momenta}
\end{equation}
where $\theta_*$ and $\phi_*$ describe the outgoing electron.
The CM energies and momentum magnitude are
\begin{equation}
\begin{aligned}
E_e^*&=\frac{s_{\rm M}+m_e^2-m_\chi^2}{2\sqrt{s_{\rm M}}},&
E_\chi^*&=\frac{s_{\rm M}+m_\chi^2-m_e^2}{2\sqrt{s_{\rm M}}},\\
p_*^2&=\frac{\lambda_{\rm K}(s_{\rm M},m_e^2,m_\chi^2)}{4s_{\rm M}},&
\lambda_{\rm K}(a,b,c)&=a^2+b^2+c^2-2ab-2ac-2bc,\\
p_*^x&=\left[
\frac{\gamma_{\rm CM}-1}{\beta_{\rm CM}^2}\beta_{\rm CM}^z p_e-\gamma_{\rm CM} E_e
\right]\beta_{\rm CM}^x,&
p_*^z&=
p_e+
\left[
\frac{\gamma_{\rm CM}-1}{\beta_{\rm CM}^2}\beta_{\rm CM}^z p_e-\gamma_{\rm CM} E_e
\right]\beta_{\rm CM}^z
\end{aligned}
\label{sup:cm-energies}
\end{equation}

By applying the inverse of the Lorentz boost defined above, the final electron energy and energy loss in the flow frame
are,
\begin{align}
E_{e,f}
&=\gamma_{\rm CM}\left(E_e^*
+\boldsymbol{\beta}_{\mathrm{CM}}\cdot\boldsymbol{p}_3^*\right)
= \gamma\left(E_e^*
+\beta^x_{\mathrm{CM}} p_*\sin\theta_*\cos\phi_*
+\beta^z_{\mathrm{CM}} p_*\cos\theta_*\right),\\
\omega&\equiv E_e-E_{e,f}=-\Delta E\,.
\label{sup:energy-transfer}
\end{align}
Positive $\omega$ corresponds to energy transferred from the electron
to DM; both signs are retained in the cooling rate.

For the interaction considered in the main text, the squared matrix
element, averaged over incident spins and summed over final spins, is~\cite{Cao:2020bwd,Li:2022dqa,Chen:2025twc}
\begin{align}
\overline{|\mathcal{M}|^{2}}
=\frac{2\epsilon^2g_{\rm em}^2g_\chi^2}{(m_{A'}^2-t_{\rm M})^2}
\Big[(s_{\rm M}-m_e^2-m_\chi^2)^2
 +(u_{\rm M}-m_e^2-m_\chi^2)^2+2t_{\rm M}(m_e^2+m_\chi^2)\Big]\,,
\label{sup:matrix-element}
\end{align}
where the remaining Mandelstam variables $t_{\rm M}=(p^*_1-p^*_3)^2$ and $u_{\rm M}=2m_e^2+2m_\chi^2-s_{\rm M}-t_{\rm M}$ are computed by using Eq.~\ref{sup:cm-energies}.
Thus $E_e$, $E_\chi$, and the incident angle $\theta$ fix the CM
kinematics, while $\theta_*$ and $\phi_*$ specify the final state. Both of them determine the function of $\omega$ in Eq.~\ref{sup:energy-transfer} and the squared matrix element in Eq.~\ref{sup:matrix-element}.

For the monoenergetic, isotropic DM distribution assumed in the main
text, write
\begin{align}
f_\chi(p,r)=\frac{n_\chi(r)}{4\pi p_\chi^2(r)}\,
 \delta\!\left(p-p_\chi(r)\right),\quad \int d^3p\,f_\chi&=n_\chi=\frac{\rho_\chi}{m_\chi}\,.
\end{align}
Here, $\rho_\chi(r)$ is the DM rest-mass density and
$v_\chi(r)=(r_g/r)^{1/2}$, $E_\chi(r)=m_\chi(1-v_\chi^2(r))^{-1/2}$, $p_\chi(r)=E_\chi (r)v_\chi(r)$. 
Therefore, the single-electron energy-loss rate defined in
Eq.~\eqref{eq:coolingreduced} can be written explicitly as
\begin{equation}
\begin{aligned}
\dot E_e(E_e,r)
&=
\frac{\rho_\chi(r)}{4\pi m_\chi}
\int d\Omega\int
(-\Delta E)\,
\frac{\overline{|\mathcal M|^2}}
     {(2E_\chi)(2E_e)}
\,d\Pi_2
\\
&=
\frac{n_\chi(r)}{8E_eE_\chi}
\int_{-1}^{1}\dd\cos\theta
\int\dd\Omega_*\,
\frac{p_*}{16\pi^2\sqrt{s_{\rm M}}}\,
\omega\,\overline{|\mathcal M|^2}.
\end{aligned}
\label{sup:single-electron-rate}
\end{equation}
In obtaining the second equality, we have integrated over the azimuthal angle of the incident DM momentum, and used  $d\Pi_2
=\frac{p_*}{16\pi^2\sqrt{s_{\rm M}}}\,d\Omega_*$, and $\Omega_*$ denotes the outgoing-electron phase space in the CM frame. We then do thermal average on the electron energy-loss rate, assuming electrons follow the Fermi-Dirac distribution characterized by temperature $T_e(r)$.

% Thermally averaging over the Maxwell--J\"uttner electron distribution
% then gives
% \begin{equation}
% \begin{aligned}
% q^{\chi e}(r)
% &=n_e(r)\int_{m_e}^{\infty}\dd E_e\,
%  W_e(E_e;T_e(r))\,\dot E_e(E_e,r),\\
% W_e(E_e;T_e)
% &=\frac{E_e\sqrt{E_e^2-m_e^2}}{m_e^2T_eK_2(m_e/T_e)}
%  e^{-E_e/T_e},
% \end{aligned}
% \label{sup:thermal-rate}
% \end{equation}
% where $K_2$ is a modified Bessel function and
% $\int_{m_e}^{\infty}\dd E_e\,W_e=1$.

\section{Temperature profiles and mediator-dependent constraints}
\label{sup:sec-results}

\begin{figure}[!htb]
\centering
\includegraphics[height=0.34\textwidth,trim={7pt 7pt 6pt 6pt},clip]{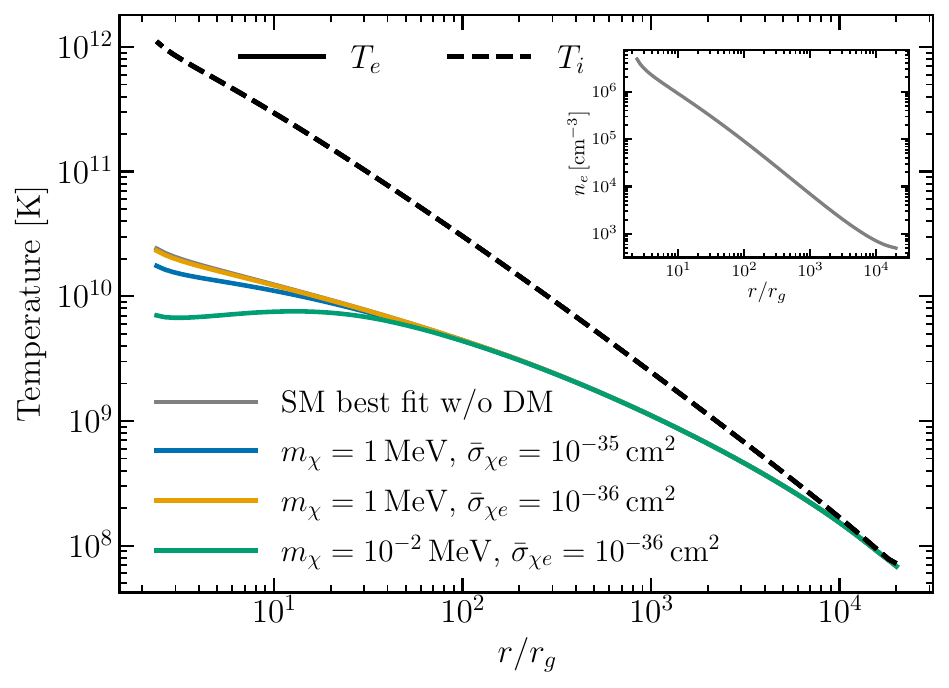}\hfill
\includegraphics[height=0.34\textwidth,trim={2pt 1pt 41pt 42pt},clip]{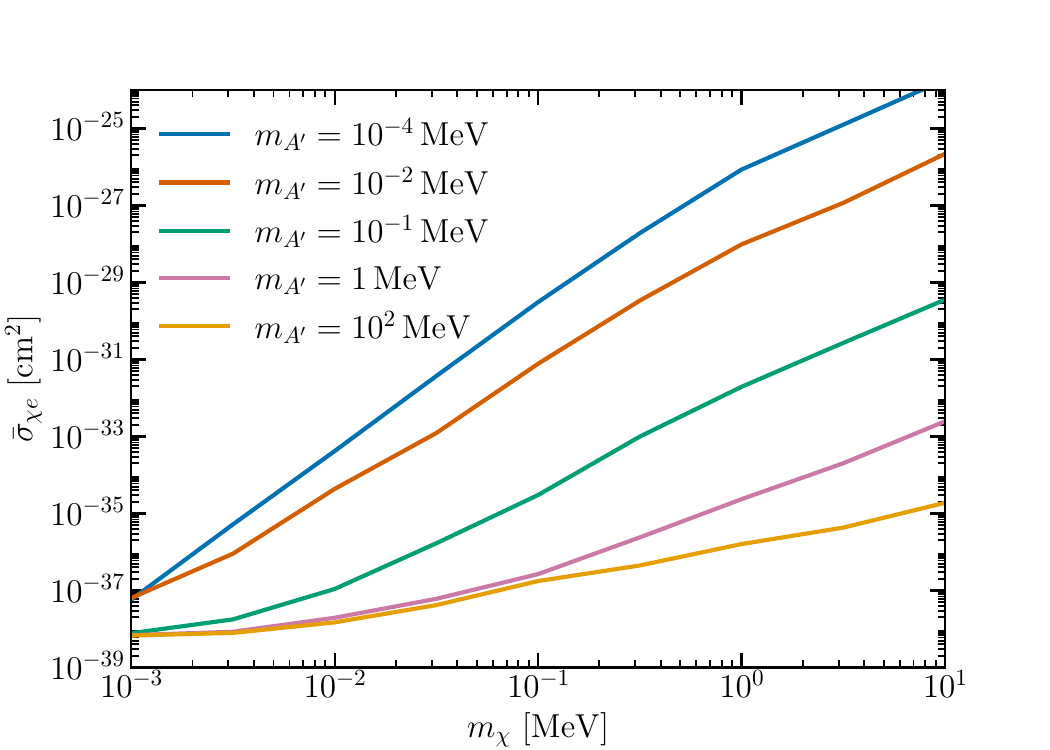}
\caption{\textit{Left: } Electron temperature $T_e$ (solid curves) and ion temperature
$T_i$ (dashed curves) as a function of the radius. Different DM masses and cross sections are chosen for a heavy mediator. 
Light-mediator benchmarks give qualitatively similar electron cooling. The inset shows the corresponding electron
number-density profile, which changes little with or without DM interaction.   \textit{Right:} 90\% C.L. upper limits on the reference DM--electron cross section $\bar\sigma_{\chi e}$ versus DM mass
$m_\chi$ for different mediator masses given in the legend.}
\label{sup:fig-temperature}
\end{figure}

Fig.~\ref{sup:fig-temperature} left shows the ion and electron temperature
profiles computed with the best-fit baseline described in
Table~\ref{sup:tab-parameters} and Fig.~\ref{sup:fig-mcmc}. 
DM cooling reduces $T_e$ in the inner flow, while $T_i$ remains nearly
unchanged.  These local temperatures enter both the
thermal average and the emission
calculation.
A lower DM mass or
a larger cross section produces stronger electron cooling. In contrast, the electron number density is tied to the ion number density through charge neutrality. DM cooling only weakly perturbs the ion pressure that supports the flow, so both number densities change little.

Fig.~\ref{sup:fig-temperature} right presents the constraints on $\bar\sigma_{\chi e}$  different mediator masses $m_{A'}$. The spread between the mediator-dependent limits is small toward the
lowest DM masses shown and grows with $m_\chi$, with the
lighter-mediator curves losing sensitivity more rapidly at larger $m_\chi$. This behavior follows from the momentum dependence of
Eq.~\eqref{sup:matrix-element}. 
For a light mediator, larger invariant transfers $-t_M$ suppress the
rate relative to the heavy-mediator.

\end{document}